\documentclass[11pt, a4paper]{article}
\usepackage{moriond,epsfig}

\def\be{\begin{equation}}
\def\ee{\end{equation}}
\def\bea{\begin{eqnarray}}
\def\eea{\end{eqnarray}}

\begin{document}
\vspace*{4cm}
\title{Searches for non-Standard-Model Higgs Bosons at the Tevatron}

\author{Greg LANDSBERG}
\address{Brown University, 182 Hope St, Providence, RI 02912, USA}
\maketitle\abstracts{ 
Search for non-Standard-Model Higgs bosons is one of the major goals of the ongoing Fermilab Tevatron run. Large data sets accumulated by the CDF and D\O\ experiments break new grounds in sensitivity. We review recent Tevatron results on searches for Higgs bosons in Minimal Supersymmetric Model in the multi $b$-jet and $\tau\tau$ final states, as well as a search for fermiophobic Higgs in the multiphoton final state.} %
\noindent


Search for the elusive Higgs bosons is among the most exciting physics topics offered by the high-luminosity Run II of the Fermilab Tevatron. In the Standard Model (SM), the Higgs boson $h^0$ is responsible for Electroweak Symmetry breaking (EWSB) and gives rise to fermion and vector gauge boson masses. However, the Higgs sector may be more complicated than the SM predicts. Generic Two-Higgs Doublet Models (2HDM) contain two complex Higgs doublets, which coupled to up and down-type quarks, respectively. This allows for eight degrees of freedom corresponding to the two complex doublet fields $H_u$ and $H_d$. The ratio of vacuum exectation values of the $H_u$  and $H_d$ doublets is traditionally denoted as $\tan\beta$, which is one of the major parameters in 2HDM. The value of $\tan\beta \approx 45$ equal to the ratio of the top and bottom quark masses (measured at the $Z$ pole) is often referred to as the ``golden'' value.

Perhaps the most successful and economical theoretical realization of 2HDM is the Minimal Supersymmetric Model (MSSM). In MSSM EWSB occurs naturally via radiative corrections. As a result, three out of the above eight degrees of freedom become longitudinal polarizations of the $W^\pm$ and $Z$ bosons, thus giving rise to their masses. The remaining five degrees of freedom correspond to four scalar Higgs bosons: $h^0$, $H^0$, and $H^\pm$, and one pseudoscalar boson, $A^0$. The SM-like $h^0$ boson is expected to be light, with the mass less than $\sim 135$ GeV, whereas the other four Higgses can be much heavier.

\begin{figure}[tbh]
\begin{center}
\psfig{figure=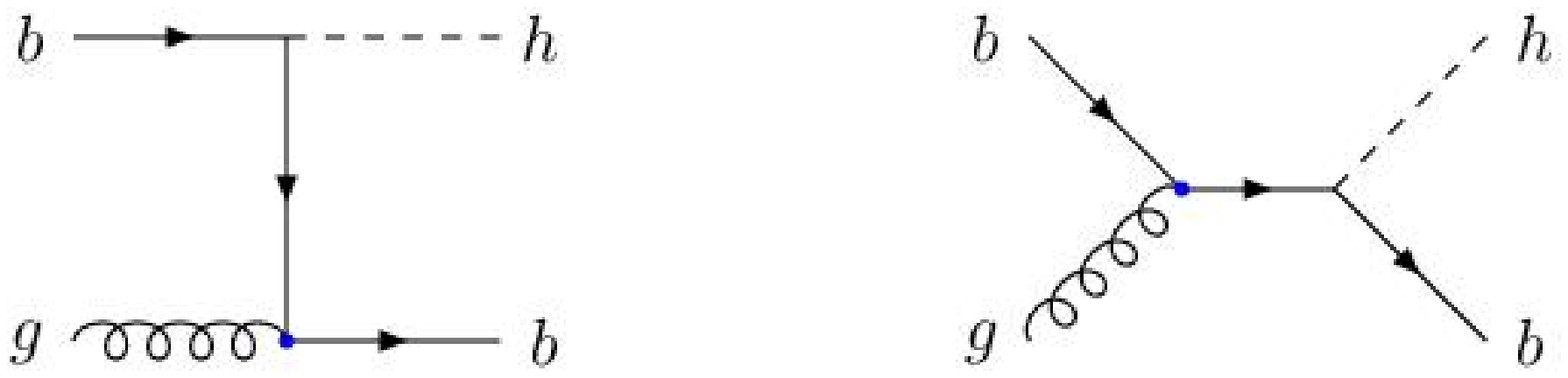,height=2.5cm} 
\psfig{figure=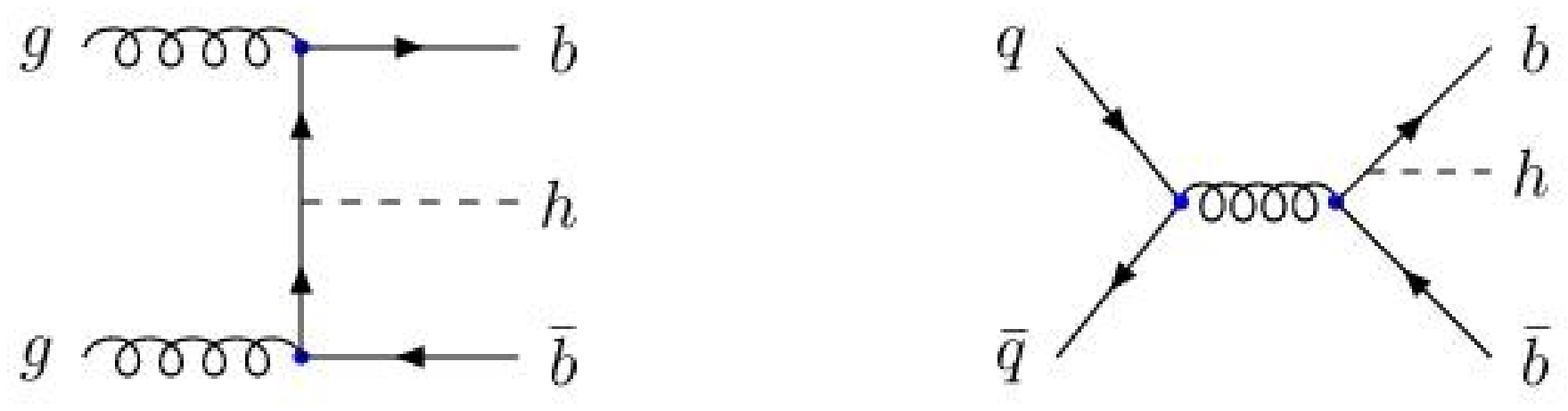,height=2.5cm}\vspace{-0.5cm}
\caption{Feynman diagrams for the MSSM Higgs production at the Tevatron.
\label{fig:diagrams}}
\end{center}
\end{figure}

At the Tevatron, the MSSM Higgs bosons $A$ and $H$ are mainly produced either via gluon fusion or radiated off b-quarks (see Fig.~\ref{fig:diagrams}). Production cross section of the $H$ and $A$ bosons is typically enhanced compared to the SM Higgs by the $\tan^2\beta$ factor, which may be quite large. This enhancement results in high sensitivity to these particles at the Tevatron, up to their masses of $\sim 200$ GeV, which covers a significant fraction of the allowed MSSM parameter space.

For large $\tan\beta$, where the Tevatron offers maximum sensitivity, the pseudoscalar Higgs $A$ is nearly degenerate in mass with either $h^0$ or $H^0$. Therefore, the final states of interest receive contribution from both the pseudoscalar and one of the scalar Higgses. In what follows we will refer to $A$ and $H$ or $h$ collectively as $\phi$ and won't distinguish between the scalar and pseudoscalar Higgses. The major decay modes of $\phi$ are $b\bar b$ ($\sim 90\%$) and $\tau\tau$ ($\sim 10\%$).


In a class of 2HDM, couplings of the light Higgs to fermions are suppressed. We refer to such case as a ``fermiophobic Higgs.'' As fermiophobic Higgs decay to $b\bar b$ is suppressed, for a sufficiently light Higgs boson the major decay channel could be $\gamma\gamma$, which proceeds via a loop diagram. D\O\ Collaboration has performed a search for such a fermiophobic Higgs $h_f$ produced in association with a charged Higgs. The charged Higgs is assumed to always decay into $h_f + W^{(*)}$, while $h_f$ in turn decays to two photons with 100\% probability. The full process is therefore $p \bar p \to h_f H^\pm \to h_f h_f W^{\pm(*)} \to 4\gamma + X$. Experimentally, it is sufficient to require at least three photons in the final state, which results in a very low background and maximizes signal efficiency. The analysis is based on $\approx 0.8$ fb$^{-1}$ of data collected with a suite of electromagnetic (EM) triggers. The following selection is used in the analysis: at least three photons with the transverse energies above 30, 20, and 15 GeV, respectively, located in the central calorimeter ($|\eta| < 1.1$) and passing standard quality requirements. This selection corresponds to five events observed in data with the expected background of $3.5 \pm 0.6$ events, dominated by direct triple-photon production ($2.73 \pm 0.55$ events) and multijet and direct photon events with jets misidentified as photons ($0.72 \pm 0.15$ events). Backgrounds are further reduced by requiring the transverse momentum of the $3\gamma$ system to exceed 25 GeV. This requirement reduces background to $1.1 \pm 0.2$ events and leaves no candidates in data. In the absence of a signal, the following limit is set on the $h_f$ production: $\sigma(h_f H^\pm) < 25.3$~fb at the 95\% confidence level (C.L.). This cross section limit can be interpreted as the mass limit on $h_f$ as a function of the charged Higgs mass and $\tan\beta$. Since associated $h_f H^\pm$ production depends on $\tan\beta$ only weakly, the mass limits are similar for small and large values of $\tan\beta$. For example, $m_h < 66$ $(80)$~GeV have been excluded at the 95\% C.L. for the charged Higgs mass $< 100$~GeV and $\tan\beta = 3$ $(30)$. The limits become less restrictive with the charged Higgs mass increase. For example, they drop to 44 (60) GeV for the charged Higgs mass $< 150$~GeV.

\begin{figure}[tbh]
\begin{center}
(a)\psfig{figure=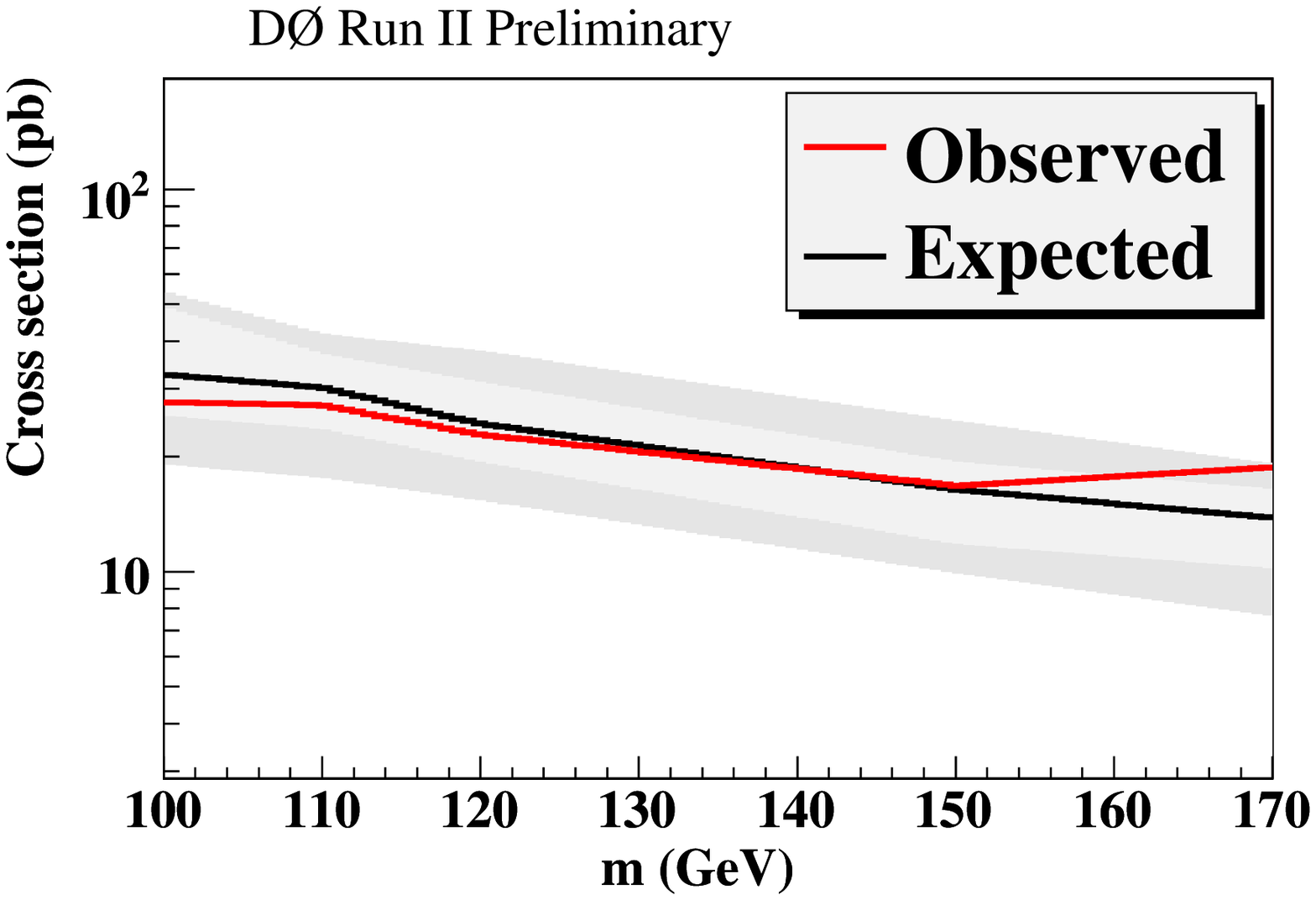,height=5.0cm}(b)
\psfig{figure=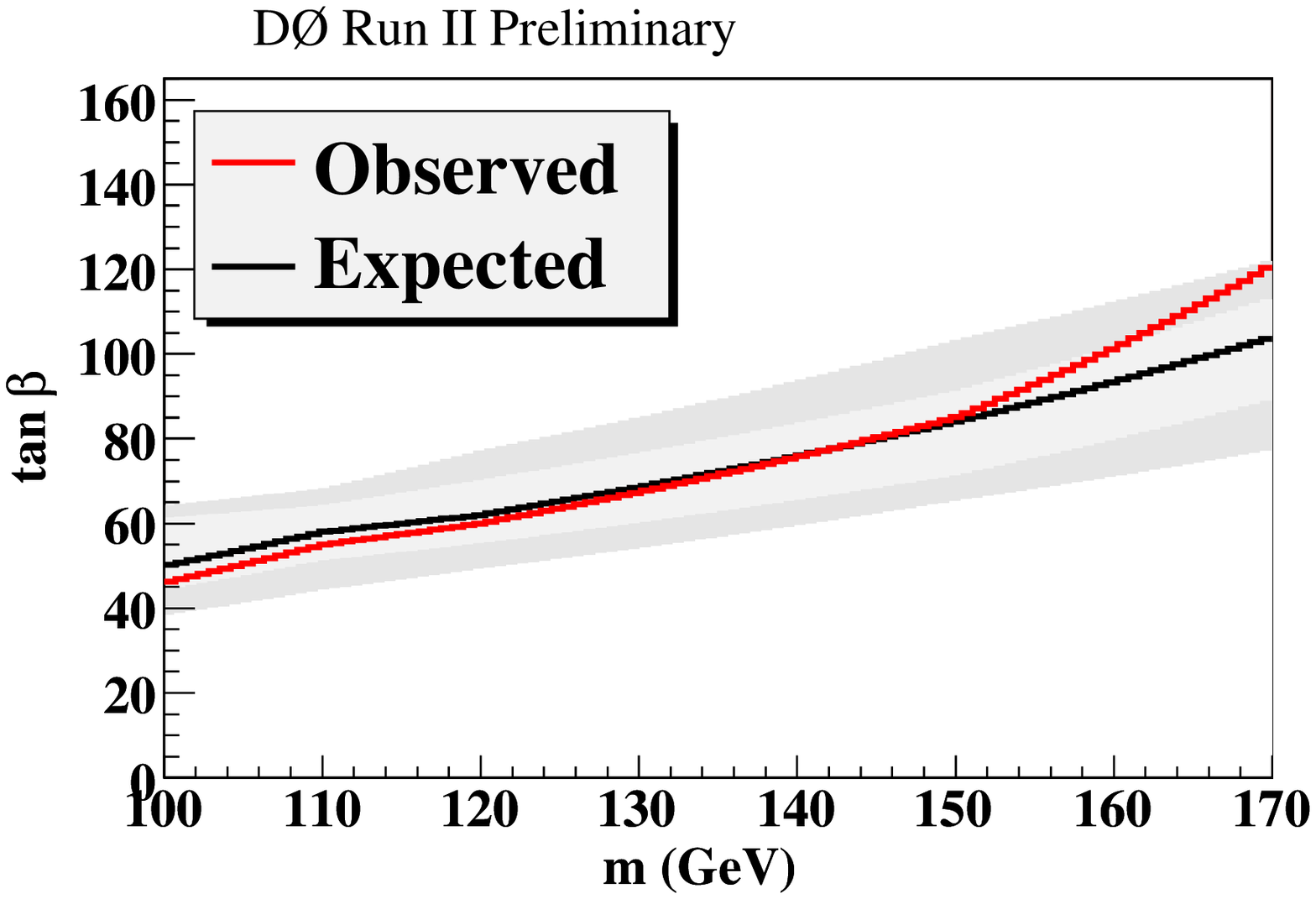,height=5.0cm}
\caption{95\% C.L. limits on (a) the MSSM Higgs production cross section and (b) $\tan\beta$ as a function of the Higgs mass. Area above the red line is excluded.
\label{fig:bbb}}
\end{center}
\end{figure}

Recently, D\O\ reported a new result of the search for $\phi$ production in association with one or two $b$-quarks, which corresponds to the 3$b$ or 4$b$ final states. An improved $b$-tagging technique based on several tagging algorithms combined via a neural net has been used, resulting in approximately 50\% higher efficiency compared to the performance of the best single tagging algorithm. The analysis is based on $\sim 0.9$ fb$^{-1}$ of data. The selection required at least three $b$-tagged jets with transverse energies above 40, 25, and 15 GeV. Signal is then searched for by looking for an excess of events in the invariant mass spectrum of the two leading jets in a mass window around the assumed $\phi$ mass. The main background comes from multijet production and is estimated from an orthogonal data set selected in a similar way to the signal sample, except for the requirement of exactly two $b$-tagged jets. Good agreeement is observed between the predicted background and data in the invariant mass spectrum of the two leading jets. Limits on Higgs production cross section as a function of its mass have been set and are shown in Fig.~\ref{fig:bbb}a. Using leading-order (LO) cross section calculations, they are interpreted as limits in the $m_\phi$-$\tan\beta$ plane as shown in Fig.~\ref{fig:bbb}b. For low $\phi$ mass the sensitivity of the search approaches the golden value for $\tan\beta$. Note that the limits on $\tan\beta$ are not directly comparable to the  published D\O\ results~\cite{d0-bbb} as the latter are based on next-to-LO cross section.

While the branching fraction of $\phi \to \tau\tau$ decay is roughly an order of magnitude less than that in the $b\bar b$ channel, the $\tau\tau$ final state is cleaner than the multijet one, thus offering competitive sensitivity to the MSSM Higgs. Both the CDF and D\O\ Collaborations have recently reported on new searches for supersymmetric Higgs in the $\tau\tau$ channel based on $\sim 1$~fb$^{-1}$ data sets. The CDF analysis requires one of the $\tau$'s to decay leptonically in the electron or muon channel, and the other $\tau$ to decay either hadronically or leptonically. In the case of both $\tau$'s decaying leptonically, one is required to decay into the electron, while the other~-- in the muon channel. The D\O\ analysis is performed in the muon-hadron channel only, but offers competitive sensitivity to the CDF analysis mainly due to more sophisticated neural-net-based $\tau$-identification, which allows for a higher signal efficiency and lower background.

\begin{figure}[t]
\begin{center}
(a)\psfig{figure=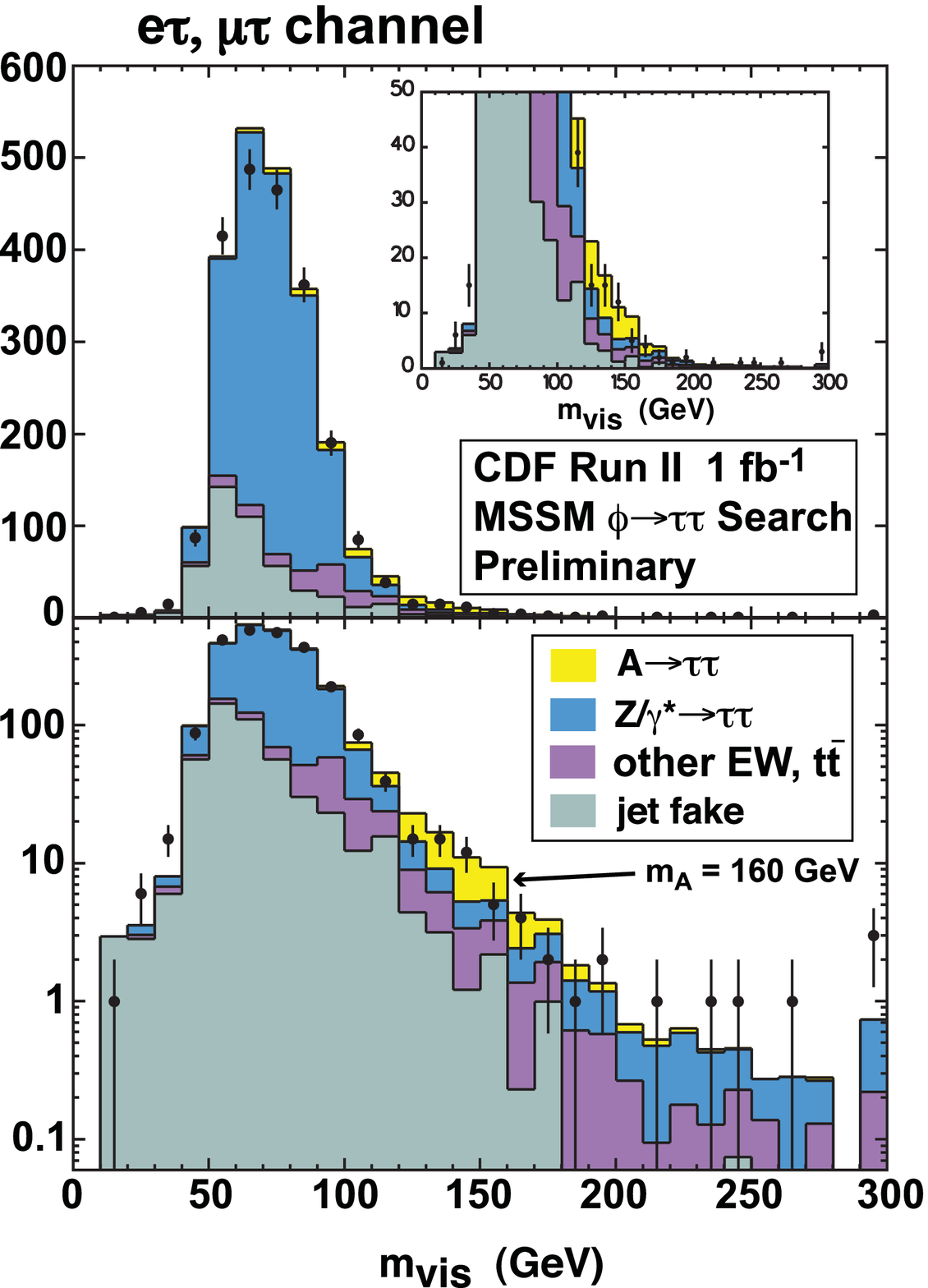,height=7.5cm}(b)
\psfig{figure=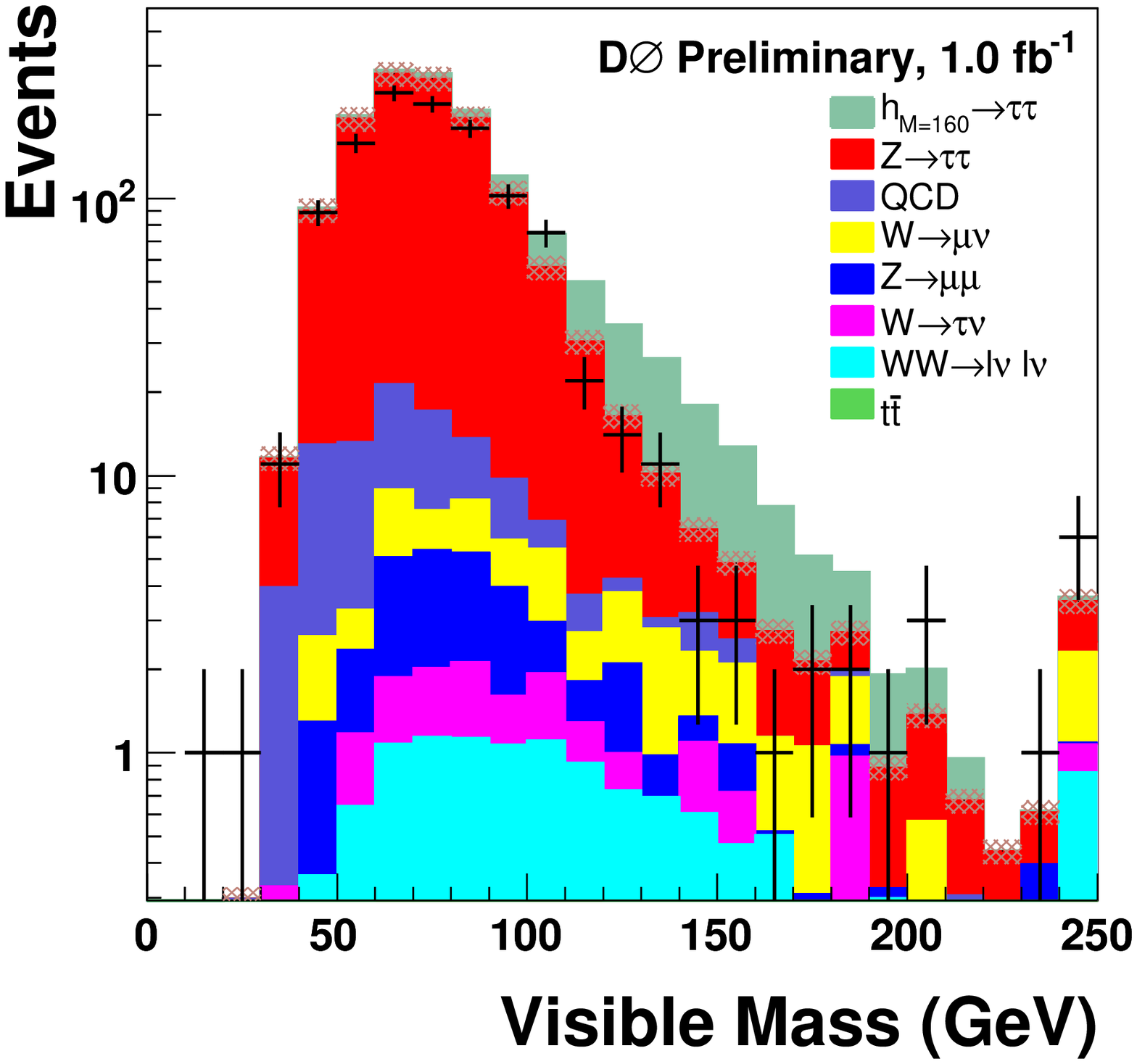,height=8.0cm}
\caption{Visible $\tau\tau$ mass spectrum in the leptonic-hadronic channel for (a) the CDF and (b) the D\O\ analysis.
\label{fig:ditaus}}
\end{center}
\end{figure}

\begin{figure}[t]
\begin{center}
(a)\psfig{figure=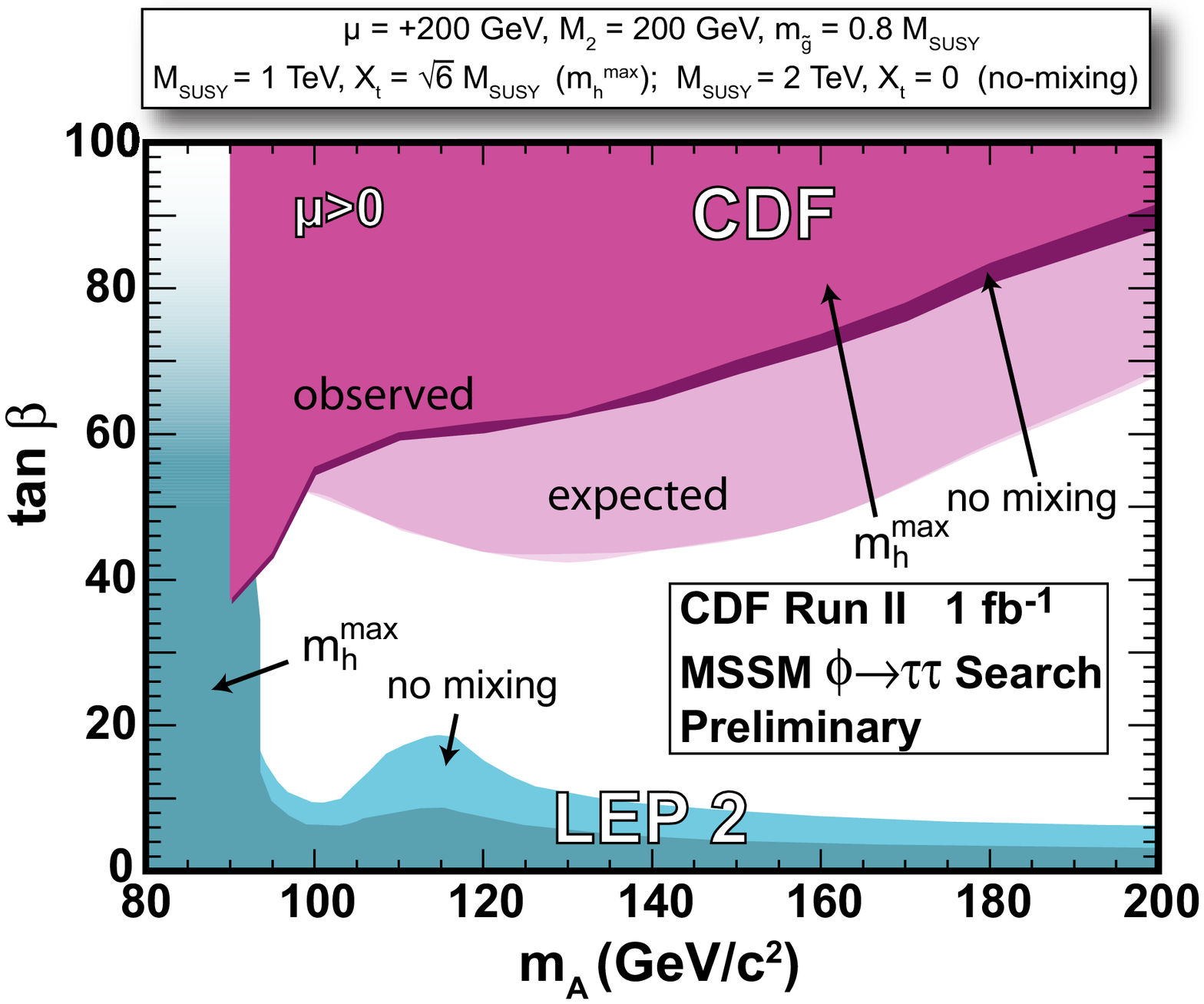,height=5.0cm}(b)
\psfig{figure=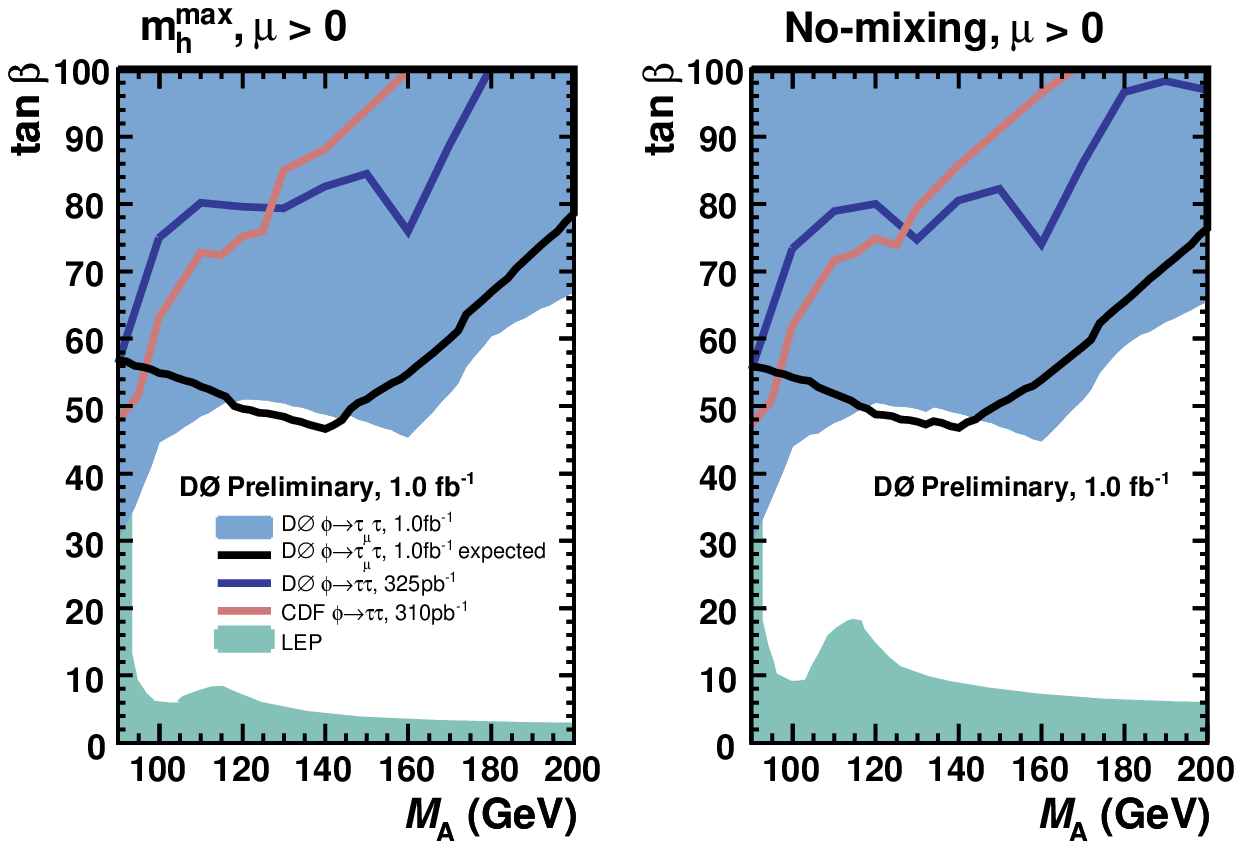,height=5.5cm}
\caption{Exclusion in the $m_\phi$-$\tan\beta$ plane for (a) the CDF and (b) the D\O\ analysis.
\label{fig:tautau}}
\end{center}
\end{figure}

While the total background predictions for the CDF analysis in all three channels agree well with the observed number of events, a slight excess of events is seen in the visible $\tau\tau$ invariant mass spectrum at the mass $\sim 160$~GeV. This excess is consistently seen in both lepton-hadron channels and is consistent with the production of a MSSM Higgs with the mass $\sim 160$ GeV and $\tan\beta \sim 50$ (see Fig.~\ref{fig:ditaus}a). However, as the significance of the excess (taking into account that it could have been observed at any mass) is less than two standard deviations, CDF still reports limits on the MSSM Higgs production in this channel, which are hurt by the observed excess (see Fig.~\ref{fig:tautau}a illustrating that the expected sensitivity is significantly better than the observed limits).

Given this exciting hint from CDF, it is very interesting to see the analogous D\O\ analysis results. Unfortunately, D\O\ actually sees a deficit of events in the region of the CDF excess (see Fig.~\ref{fig:ditaus}b, where a 160 GeV Higgs signal is indicated with shading). Translated into limits on the MSSM Higgs, the D\O\ analysis excludes the MSSM Higgs with the mass of 160 GeV for $\tan\beta > 40$, thus largely disfavoring an interpretation of CDF excess with the MSSM Higgs production.

More data being collected in the ongoing Tevatron run will allow both Collaborations to extend the reach for non-SM Higgs searches even further and either discover the Higgs or set stringent limits on the MSSM parameters.

\section*{Acknowledgments}

I'd like to thank the Moriond QCD conference organizers for a kind invitation and hospitality. Many thanks to my CDF and D\O\ colleagues for providing me with the material for this talk and for helpful discussions. Special thanks to Gregorio Bernardi, John Conway, Andy Haas, Mark Owen, and Stefan Soldner-Rembold. This work is partially supported by the U.S.~Department of Energy under Grant No. DE-FG02-91ER40688 and by the National Science Foundation under the CAREER Award PHY-0239367.

\end{document}